\documentclass{egpubl}
\JournalPaper         
\electronicVersion 

\ifpdf \usepackage[pdftex]{graphicx} \pdfcompresslevel=9
\else \usepackage[dvips]{graphicx} \fi

\PrintedOrElectronic

\usepackage{t1enc,dfadobe}
\usepackage{cite}
\usepackage{egweblnk}

\usepackage{overpic}
\usepackage{amssymb}
\usepackage{amsmath}
\usepackage{graphicx}
\usepackage{wrapfig}
\usepackage{float} 
\usepackage{caption} 
\usepackage{enumitem} 
\usepackage{microtype}
\usepackage{multirow}
\DeclareMathOperator*{\argmin}{arg\,min}

\newcommand{\myfigurename}{}

\newcommand{\Figure}[1]{Figure~\ref{fig:#1}}
\newcommand{\eq}[1]{(\ref{eq:#1})}

\newcommand{\Equation}[1]{Equation~\ref{eq:#1}}

\newcommand{\Section}[1]{Section~\ref{sec:#1}}

\newcommand{\Table}[1]{Table~\ref{tab:#1}}

\providecommand{\citename}[2]{{#1~et~al.~\cite{#2}}}

\providecommand{\citet}[1]{\todo{Blah et al.~\cite{#1}}}

\newcommand{\brief}[1]{} 
\newcommand{\ignore}[1]{}

\usepackage{currfile}

\newcommand{\vertical}[1]{{\rotatebox{90}{#1}}}

\usepackage{comment}

\renewcommand{\paragraph}[1]{{\textbf{#1.}}}

\usepackage{amsthm}
\newtheorem{defn}{Definition}[section]
\newenvironment{definition}[1][Definition]{%
\vspace{-2\parskip}
\begin{quote}\begin{defn}} {\end{defn}\end{quote}}
	
\usepackage{pifont}
\providecommand{\circledone}{\ding{172}}
\providecommand{\circledtwo}{\ding{173}}

\let\c\undefined
\let\r\undefined
\newcommand{\r}{r}
\newcommand{\c}{\mathbf{c}}
\newcommand{\x}{\mathbf{x}}
\newcommand{\matradii}{r}
\newcommand{\matcenter}{\mathbf{c}}
\newcommand{\kernel}{\varphi}

\newcommand{\object}{\mathcal{O}}
\newcommand{\surface}{\partial\mathcal{O}}
\newcommand{\MAT}{\text{MAT}}
\newcommand{\axis}{\mathcal{M}}
\newcommand{\radfun}{\mathcal{R}}

\newcommand{\normal}{\mathbf{n}}
\newcommand{\point}{\mathbf{p}}
\newcommand{\footpoint}{\mathbf{f}}
\newcommand{\sphere}{\mathbf{s}}
\newcommand{\bubble}{sphere-shrinking}
\newcommand{\groundtruth}{\tilde}

\newcommand{\ramp}{\mathcal{R}}
\newcommand{\heaviside}{\mathcal{H}}
\newcommand{\points}{\mathcal{P}}
\newcommand{\sdf}{\Phi}

\newcommand{\support}{h_\text{support}}
\newcommand{\blend}{h_\text{blend}}

\title[LSMAT: Least Squares Medial Axis Transform]{LSMAT \\ Least Squares Medial Axis Transform}
\author[Rebain et al.]{
\parbox{\textwidth}{\centering Daniel Rebain$^{1,2}$, Baptiste Angles$^{1,2}$, Julien Valentin$^{2}$, Nicholas Vining$^{2}$, Jiju Peethambaran$^{1}$, Shahram Izadi$^{2}$, Andrea Tagliasacchi$^{1,2,3}$}\\{
\parbox{\textwidth}{
\centering 
$^1$Department of Computer Science, University of Victoria, Victoria, BC, Canada \\
$^2$Google LLC, Mountain View, CA, USA \\
$^3$David R. Cheriton School of Computer Science, University of Waterloo, Waterloo, ON, Canada \\
}}}

\begin{document}
\teaser{
\includegraphics[width=\linewidth]{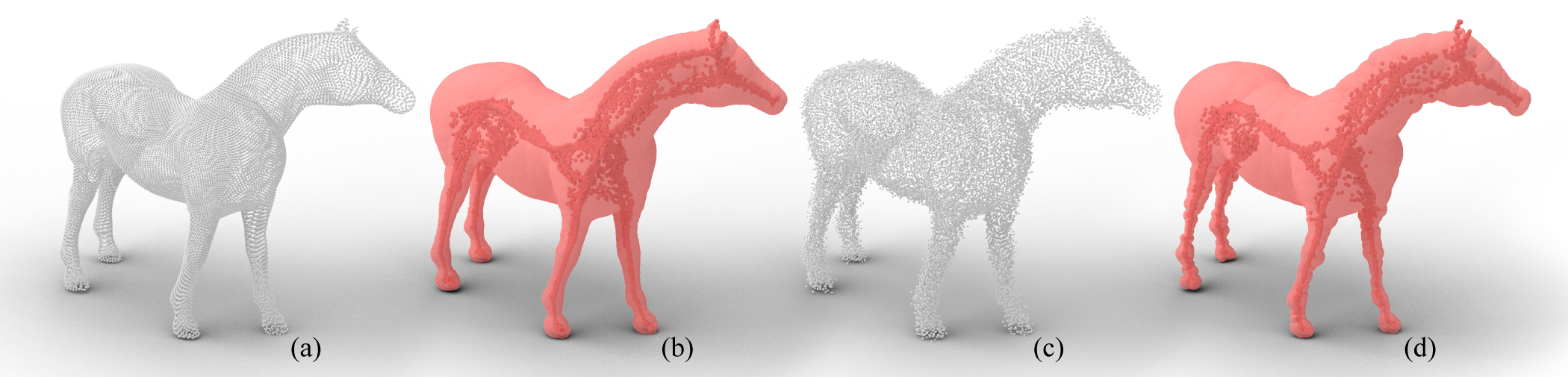}
\centering
\caption{
The LSMAT is a novel efficient least-squares formulation of the medial axis transform that operates on unorganized oriented point sets.
(a, b) Regardless of input noise, the resulting medial representation is \emph{stable}.
(c, d) Its least-squares nature allows it to operate in the presence of heavy noise where most approaches would fail. 
We visualize the oriented point cloud with oriented splats, draw the union of medial spheres in light red, and their corresponding centers in dark red.
}
\label{fig:teaser}
}

\maketitle

\begin{abstract}
The medial axis transform has applications in numerous fields including visualization, computer graphics, and computer vision. Unfortunately, traditional medial axis transformations are usually brittle in the presence of outliers, perturbations and/or noise along the boundary of objects.
To overcome this limitation, we introduce a new formulation of the medial axis transform which is naturally robust in the presence of these artifacts.
Unlike previous work which has approached the medial axis from a computational geometry angle, we consider it from a numerical optimization perspective. In this work, we follow the definition of the medial axis transform as ``the set of maximally inscribed spheres". We show how this definition can be formulated as a least squares relaxation where the transform is obtained by minimizing a continuous optimization problem. The proposed approach is inherently parallelizable by performing independant optimization of each sphere using Gauss-Newton, and its least-squares form allows it to be significantly more robust compared to traditional computational geometry approaches. Extensive experiments on 2D and 3D objects demonstrate that our method provides superior results to the state of the art on both synthetic and real-data.
\keywords
\keyword{medial axis transform}
\keyword{optimization}
\keyword{least squares}
\endkeywords
\ccsdesc[500]{Computing methodologies~Point-based models}
\ccsdesc[300]{Computing methodologies~Volumetric models}
\ccsdesc[100]{Computing methodologies~Shape analysis}
\printccsdesc
\end{abstract}

\begin{figure*}[t]
\centering
\begin{overpic} 
[width=\linewidth]
{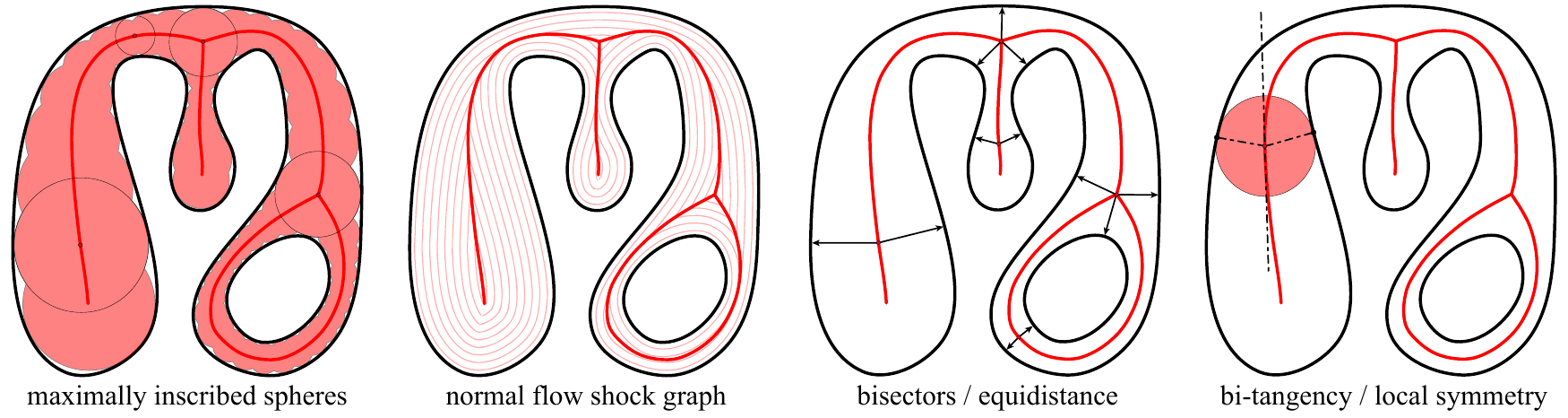}
\myfigurename{}
\end{overpic}
\caption{
Visualization of four alternative definitions of medial axis; base image courtesy of~\cite{skelstar}.
}
\label{fig:definitions}
\end{figure*}
\section{Introduction}
A medial representation of an object encodes its solid geometry as the union of a collection of spheres of different radii and origins; see~\Figure{definitions}.
While volumetric or surface mesh representations are more commonly used in computer graphics and computer vision, since its introduction by~\citename{Blum}{blum}, medial representations have found applications in many 2D/3D geometric problems such as animation~\cite{pinocchio}, fabrication~\cite{musialski2016fabrication}, image processing~\cite{amat}, shape analysis~\cite{shapediam}, and real-time tracking~\cite{hmodel}.
The process of converting an input model into a medial representation is generally referred to as the Medial Axis Transform (MAT), and the collection of origins of the spheres in a medial representation form a {\em medial skeleton}.

\paragraph{Stability}
A common pitfall of medial axis methods is their sensitivity to noise: in a traditional medial representation, new branches of the medial skeleton may form at all negative local curvature extrema.
If our shape is represented via a piecewise linear boundary, such as a polygonal mesh in 3D, spurious branches form when the surface exhibits even minor levels of noise; see \cite[Fig.9]{skelstar}.
These issues are typically resolved by resorting to postprocessing, in which sets of spheres are filtered out according to various engineered criteria.
We present a new algorithm for computing the medial axis transform of an oriented point set which is naturally capable of handling noise, outliers, and other artifacts that create characteristic problems for traditional methods.

\paragraph{Definitions}
As summarized in \Figure{definitions} and \cite{skelstar}, the medial axis transform may be defined in a few ways, with each definition producing equally valid and useful 
representations. In practice, each definition of the medial representation leads to a different way to compute the medial axis. For example, the \emph{normal flow} variant leads to the commonly employed voxel thinning algorithms~\cite{saha2016survey}, while the \emph{equidistance} definition leads to techniques leveraging Voronoi 
diagrams~\cite{brandt92}. In this paper, we build over what is likely the most well known definition of the medial axis transform:
\begin{definition}
The Medial Axis Transform $\MAT(\object)$ of $\object$ is the set of centers $\axis$ and corresponding radii $\radfun$ of all maximally inscribed circles/spheres in $\object$.
\label{def:mat}
\end{definition}
While previous work such as \citename{Ma}{bubble} has proposed methods to construct medial axis representations using this definition, it is interesting to note that they have treated the problem from a \emph{combinatorial geometry} standpoint. Instead, we consider the medial axis transform from a \emph{numerical geometry} perspective, where the medial axis is given by the solution of an optimization problem.
We achieve this by expressing the concepts of \emph{maximality} and \emph{inscription} from Def.~\ref{def:mat} in a least squares form.
The robustness of the approach to imperfections arises from the fact that least squares optimization attempts to find an approximate, rather than exact, solution to the given problem.
We are motivated in our approach by considering a least-squares problem as a maximum likelihood estimate of a function in the presence of noise obeying a Gaussian probability distribution.

\paragraph{Method outline}
Our method takes an \emph{oriented} point cloud as input, and produces an unconnected medial point cloud as output by minimizing a combination of a maximality energy, and an inscription energy. 
Our key technical challenge is in formulating these energies correctly; our maximality energy is designed to have a constant magnitude, ensuring that the optimization energy of each medial sphere is constant regardless of its radius, and our inscription energy is based around a locally supported approximation of the signed distance function of the point cloud. 
In order to prevent spheres from sliding towards local maxima of local shape thickness, we introduce a pinning constraint as a quadratic barrier energy.
The resulting optimization problem is quadratic, with differentiable but non-linear energy terms, and we solve it with an iterative Gauss-Newton solver.

\paragraph{Contributions and Evaluation}
Our main contributions are a novel interpretation of a problem that is classically solved by standard computational geometry as a numerical geometry problem, and a novel algorithm for computing the medial axis transform of an oriented point set that inherently handles imperfections in the input. We present a number of qualitative results throughout the paper for both 2D and 3D oriented point clouds. Finally, we also present side-by-side quantitative evaluations of the proposed algorithm against several state-of-the-art methods.

\section{Related Work}
\label{sec:related}
Many techniques exist to compute the medial axis transformation, as detailed in a recent survey~\cite{skelstar}.
We focus our discussion around two central aspects: methods that operate on \emph{surface} 2D/3D representations (e.g. point clouds and 
triangular meshes), and methods that attempt to resolve the \emph{instability} of medial axis via post processing.

\subsection{Medial axis computation of sampled surfaces}
\label{sec:voronoi}
\begin{figure}[t]
\centering
\begin{overpic} 
[width=\linewidth]
{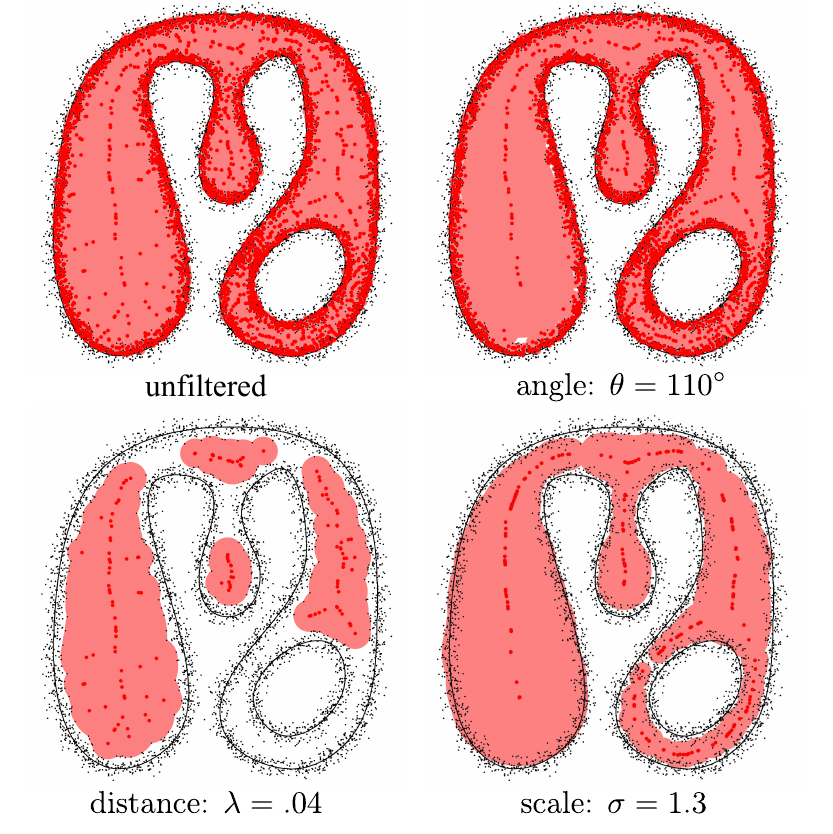}
\myfigurename{}
\end{overpic}
\caption{
Effectiveness of standard filtering Voronoi methods with high level of noise.
As expected, neither angle nor distance methods were capable of filtering out all noise for any selected threshold value.
Leveraging global information, the scale axis is able to deal with noise, but with a significant computational burden.
Further, note that we provide these technique the ground truth inside/outside labeling, which is \emph{not available} in our context.
}
\label{fig:stability}
\end{figure}

Assuming the surface is \emph{sampled} by a sufficiently dense set of points, the Voronoi diagram can be used to compute the medial axis transform with relative ease.
For a closed boundary in 2D, any interior Voronoi vertex, as well as Voronoi edges connecting interior vertices, approximate the medial axis with a convergence guarantee~\cite{brandt92}.
Unfortunately, \citename{Amenta}{amenta99} has shown how this property does not hold in 3D due to \emph{sliver tetrahedra}.
Thankfully, as we increase the sampling rate, the Voronoi poles (see definition in \cite{amenta99}) do converge to the medial axis, allowing the use of 3D Voronoi diagrams for the task at hand.
These methods suffer two fundamental shortcomings:
\circledone{} they are \emph{global} and optimize the entire set of centers at the same time, and 
\circledtwo{} the extracted axis is heavily susceptible to even minor levels of noise; see \Figure{stability}.

\begin{figure}[t]
\centering
\begin{overpic} 
[width=\linewidth]
{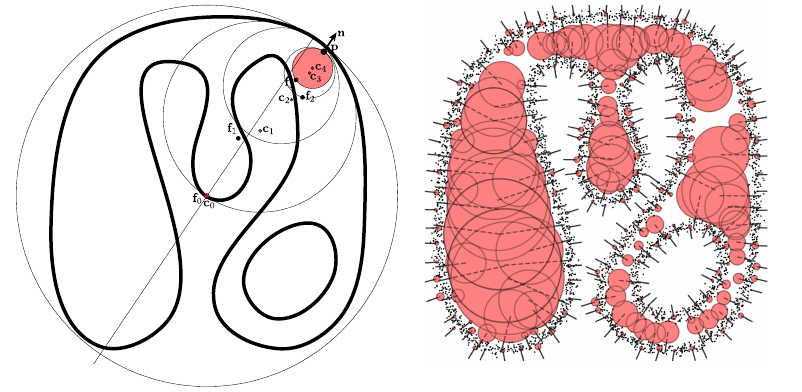}
\myfigurename{}
\end{overpic}
\caption{
(left) The \bubble{} of \citename{Ma}{bubble}, and (right) its resulting maximal spheres on a noisy point set.
}
\label{fig:bubble}
\end{figure}

\paragraph{Sphere-at-a-point methods}
Leveraging an \emph{oriented} sampling, \citename{Ma}{bubble} proposed an alternative way to compute the transform by marrying the maximally inscribed definition of \Figure{definitions}(a) to the bi-tangency of \Figure{definitions}(d). Unlike Voronoi methods, which consider the entire point set at once, their method can compute a maximal sphere at a point in \emph{isolation}, leading to extremely efficient GPU implementations~\cite{jalba_pami13}.
These methods represent the \emph{most efficient} way of computing the medial axis, but, similarly to Voronoi methods, they suffer stability issues; see \Figure{bubble}.
Another method for computing a local approximation to the axis was proposed by \citename{Shapira}{shapediam} via casting rays in a cone oriented along the anti-normal.
The distance between the point and the intersection with the surface is aggregated by a robust function (e.g. median) to estimate the \emph{shape diameter function}.
While this approximation has found widespread use as a shape descriptor, the radius estimate suffer of \emph{bias}, and the algorithm does not generalize to point clouds.

\paragraph{Shape approximation methods}
\label{sec:shapeapprox}
Recently, a new class of techniques has been proposed which attempt to \emph{approximate} watertight surfaces via \emph{Sphere~Meshes} -- linearly swept spherical primitives~\cite{spheremesh, anispheremesh}.
This is achieved via local mesh decimation relying on iterative edge collapses, where \emph{spherical quadrics} are employed in place of the traditional quadric metrics~\cite{qslim}.
In many cases the produced model \emph{resembles} a medial axis.
When executed on 3D data the result can contain tetrahedra, however, the medial axis of a 3D shape is known to consist only of points, curves, and surfaces.
Addressing these concerns, the \emph{Medial~Meshes} work by \citename{Sun}{medialmesh} extended sphere~meshes to decimate a medial axis mesh, and discard unstable branches. Marrying sphere~meshes to medial~meshes, \citename{Li}{qmat} followed up this work and proposed QMAT, a more computationally efficient version based on spherical quadrics.
While these techniques, in juxtaposition to our local method, are global, they can only cope with minor levels of noise:~\emph{``with very noisy input, however, the simplified medial mesh is not a stable representation''}; see \cite[Fig.9]{medialmesh}.
The follow-up work by~\citename{Li}{qmat} performs slightly better as it can optimize for sphere centers, but our algorithm can still cope with noise that is \emph{one order of magnitude} larger; see~\cite[Fig.16]{qmat}.

\subsection{Instability and filtering techniques}
Techniques that do not attempt to produce an \emph{approximation} suffer from instability when computing the medial axis.
Filtering techniques attempt to \emph{remove} portions of the medial axis that do not contribute significantly to the geometry \emph{reconstructed} as the union of medial spheres.
As shown in \cite[Fig.2]{scaleaxis3d}, this works fairly well for inputs with little or no noise.
Conversely, with large noise (and/or outliers), these methods become mostly inappropriate; see~\Figure{stability}.

\paragraph{Angle filtering -- $\theta$-medial axis}
One way to identify the significance of a point is the largest angle formed by the center of the corresponding maximal sphere and two of 
its tangent points on the shape boundary. The $\theta$-medial axis of~\cite{foskey}, filters out balls associated with low aperture angle. 
While this filtering can disconnect portions of the medial axis, see \Figure{stability}(c), the issue can be avoided by 
homotopy-preserving pruning~\cite{attali}. 

\paragraph{Distance filtering -- $\lambda$-medial axis}
Another metric for filtering discards a sphere whenever its tangent points lie on the surface below a certain 
distance~\cite{lambda,powercrust}. As \Figure{stability}(d) illustrates, this results in a loss of features even before all noise has been 
removed. This shortcoming makes these solutions inappropriate whenever the input shape contains structures at different scales.

\paragraph{Scale filtering -- $\sigma$-medial axis}
The scale axis transform introduced by \citename{Giesen}{scaleaxis} and extended to 3D by \citename{Miklos}{scaleaxis3d} are methods built over the \emph{maximality} property of the medial axis.
First medial spheres are scaled by a given $\sigma>1$ value, and spheres that are no longer maximal (i.e. contained in another sphere) get discarded.
The medial axis of the scaled union of balls is then computed, and its spheres unscaled by a factor $1/\sigma$.
This method is \emph{global} as it processes the whole point set at once, and computes multiple Voronoi diagrams in the process, resulting in reduced computational efficiency -- e.g. $\approx 2~min.$ for a mesh with $\approx 100k$ vertices, see~\cite[Tab.1]{scaleaxis3d}.

\begin{figure*}[t]
\centering
\begin{overpic} 
[width=\linewidth]
{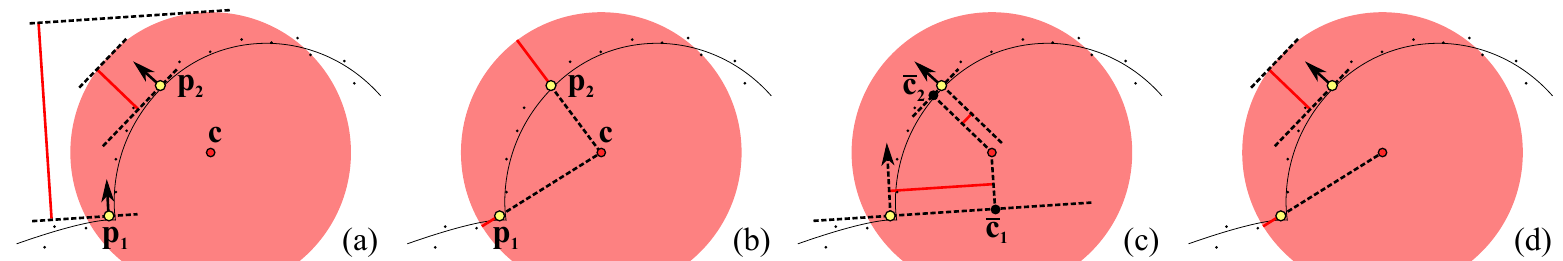}
\myfigurename{}
\end{overpic}
\caption{
(a) 
The function $ \sdf_\text{halfplane} $ uses point normals to give a signed distance value that penalizes spheres which move outside the shape, but may give nonsensical values for points belonging to parts of the surface that are not well approximated by the sphere. 
(b) 
The function $ \sdf_\text{point} $ does not suffer from this issue but does not enforce that the sphere should respect the orientation of the surface. 
(c) 
We mix between the two distances independently for each point using the distance between the point and the projection of the sphere center onto the plane defined by the point. 
(d) 
This allows us to use the sided distance only for points where it makes sense, and use the unsigned point-to-plane function otherwise.
}
\label{fig:inscription}
\end{figure*}

\section{Technical details}
\label{sec:algorithm}
In \Section{bubble}, we start by reviewing the highly relevant method of \citename{Ma}{bubble}, which computes the medial axis by searching for maximal spheres via \emph{local} analysis.
We then formulate our numerical optimization to compute medial spheres in \Section{optcore}.

\paragraph{Notation} 
We are given in input an \emph{oriented} point cloud $\{(\point_n,\normal_n)\}_{n=1 \dots N}$ drawn from the solid object $\object$ with watertight surface/boundary $\partial\object$.
The cloud is affected by noise with standard deviation $\sigma_\point$.
We give all numerical values of $\sigma_\point$ and other parameters which represent distance values as \textit{percentages} relative to the diagonal of the bounding box of the input shape.
With $\mathcal{M}$ we refer to the \emph{internal} medial axis of $\object$; see \cite[Fig.2]{skelstar}.
With $\sphere=(\c,\r)$ we indicate a sphere centered at $\c$ of radius $\r$, and with $t$ the index of the solver iteration.


\subsection{The sphere-shrinking algorithm -- \Figure{bubble}}
\label{sec:bubble}
The algorithm proposed by \citename{Ma}{bubble} is an iterative method that shrinks an initially large sphere until it satisfies the medial axis properties.
Assuming $\point$ is a point on the boundary, this process leverages two properties: 
\circledone{} that the sphere passing through $\point$ should be empty; 
and \circledtwo{} that $\point-\matcenter$ for a smooth input curve is parallel to the contact point normal $\normal$.
The \bubble{} algorithm is initialized with a large sphere, passing through $\point$, containing the entire point set, and whose center $\matcenter$ lies along the ray $(\point, -\normal)$.
Whenever the distance to the closest point $\footpoint\in\partial\object$ from $\matcenter$ is less than $\matradii$ (i.e. the sphere is not empty), the center $\matcenter$ is updated by computing the sphere tangent to $(\point,\normal)$ and passing through $\footpoint$; see \Figure{bubble}.
Finally, the loop terminates when the sphere radius change across iterations is below numerical precision.
This algorithm is highly efficient as each sample can be processed completely independently from the others, but as it treats all points as hard constraints in a combinatorial manner it does not cope well with noisy inputs.
Our formulation borrows from this one, but generalizes it by \circledone{} reformulating it into a \emph{continuous} optimization problem, and \circledtwo{} making it more \emph{robust} to noise and outliers.


\subsection{Optimizing for maximally inscribed spheres}
\label{sec:optcore}
We define a least-squares optimization capable of generating \emph{maximally inscribed spheres} given an \emph{oriented} point cloud $\points$. 
Our optimization energy for a sphere $\sphere$ consists of the combination of two terms:
\newcommand{\maximal}{\text{maximal}}
\newcommand{\wmaximal}{1}
\newcommand{\inscribed}{\text{inscribed}}
\newcommand{\winscribed}{2}
\begin{equation}
E_\text{medial} = 
\omega_\wmaximal\: E_\text{maximal} \: + \:   
\omega_\winscribed \: E_\text{inscribed}
\label{eq:maxball}
\end{equation}
We will now proceed to describe these terms in detail, and then provide comparison against alternative formulations in \Section{eval_maximal}.
 
\paragraph{Maximality}
The maximality energy creates a constant positive pressure term that tends to increase the sphere size at each iteration. 
By design, we define this energy to have a \emph{constant magnitude}, that is, a contribution to the optimization energy that is constant regardless of the radius of the given medial sphere.
This is achieved by considering the radius at the previous optimization iteration with a constant offset $\epsilon$:
\begin{equation}
E_\maximal = || \matradii - (\matradii^{t-1} + \epsilon) ||_2^2
\label{eq:maximal}
\end{equation}

\begin{figure*}[t]
\centering
\begin{overpic} 
[width=\linewidth]
{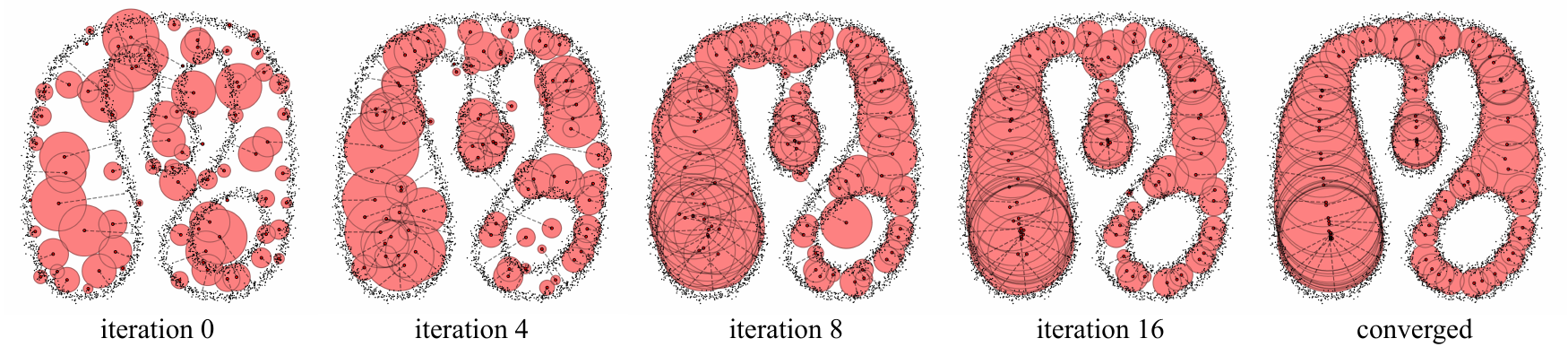}
\myfigurename{}
\end{overpic}
\caption{
Iterations of LSMAT optimization on an input noisy point set.
We randomly initialize sphere centers and radii, and demonstrate the excellent convergence properties of our optimization.
Here the center-pin correspondence is marked by the dotted line.
Notice how although some spheres are initially outside, the optimization pushes them into the interior of the point set.
}
\label{fig:iterations}
\end{figure*}

\paragraph{Inscription}
Consider a signed distance function (SDF) $\sdf(\x)$, where $\sdf(\x)<0$ for an $\x$ inside the shape.
If the expression of this function were available to us, an inscription constraint could be easily expressed by the inequality $\sdf(\c)<-r$, which we can convert in our least squares formalism as:
\begin{equation}
E_\inscribed = \ramp( r+\sdf(\c) )^2
\label{eq:inscribed_exact}
\end{equation}
where the ramp function $\ramp(x)=max(x,0)$ only penalizes a sphere when it \emph{violates} an inscription constraint.
Unfortunately, an SDF function for our oriented point set $\points$ is not readily available without computing a full surface reconstruction of the point cloud~\cite{reconstar}.
However, the family of moving least square (MLS) methods are capable of building locally supported approximations in a neighborhood of $\surface$ -- that is, as $\sdf \rightarrow 0$.
For example, the MLS formulation by Kolluri~\cite{kolluri_talg08} approximates $\sdf$ as a weighted sum of locally supported point-to-plane functions:
\begin{equation}
\sdf(\x) = \frac
{\sum_n \kernel_n(\x, h) \: \normal_n \cdot (\x-\point_n)}
{\sum_n \kernel_n(\x, h)}
\label{eq:kolluri}
\end{equation}
where in our case $\kernel_n(\x, h)=\kernel(\|\x-\point_n\|_2, h)$ and $\kernel(x, h)$ is a smoothly decaying radial basis function~\cite{apss} with compact support $[0, h]$:
\begin{equation}
\kernel(x, h) = b \left( \frac{x}{h} \right);
\quad
b(x) = 
\begin{cases} 
(1-x^2)^4& x<1 \\ 
0 & x\geq1
\end{cases}
\label{eq:kernel}
\end{equation}
Note that as \Equation{inscribed_exact} evaluates the function $\sdf$ at the medial center, which could be potentially far from $\surface$, this representation is not immediately appropriate.
However, in the context of registration, several works have shown how a quadratic approximation of $\sdf^2$ can be built by appropriately blending \emph{point-to-point} with \emph{point-to-plane} distance functions~\cite{sdfgeometry,mitra2004registration}.
In more detail, point-to-point distances are a good approximation of $\sdf^2$ in the \emph{far-field} (i.e. for $\sdf^2 \gg 0$), while point-to-plane distances are more suitable in the \emph{near-field} (i.e. for $\sdf^2 \rightarrow 0$).
Let us first define $\bar\c_n= \c^{t-1} - \normal_n(\c^{t-1}-\point_n)\cdot\normal_n$, the projection of the center on the hyperplane of point $\point_n$. We can use this to interpolate between the two metrics:
\begin{align}
\sdf_\text{blend}(\sphere)^2 =
\text{mix}(\sdf_\text{plane}(\sphere)^2, \sdf_\text{point}(\sphere)^2, \kernel_n(\bar\c_n, \blend))
\end{align}
where $\sdf_\text{plane}(\sphere) = \ramp(\r - (\point_n-\c)\cdot \normal_n)$ represents the distance from the sphere to a plane, $\sdf_\text{point}(\sphere) = \ramp(\r-\|\point_n-\c\|_2)$ is the euclidean distance from the sphere to a point, and $\text{mix}(a, b, x) = xa + (1-x)b$ is the linear interpolation operator; see \Figure{inscription}.
Based on the geometry of $\sdf_\text{blend}^2$, we can then re-formulate our inscription with respect to each oriented point $\point_n$.
Analogously to \Equation{kolluri}, this energy is accumulated over all points in $\points$:
\begin{align}
E_\inscribed \approx \sum_n
\underbrace{\kernel(\ramp(||\c^{t-1} - \point_n|| - \r^{t-1}), \support)}_{\neq 0 \text{ for a subset of the N points}}
\sdf_\text{blend}(\sphere)^2
\label{eq:inscribed}
\end{align}
The parameter $\blend$ defines the scale used for blending between distance types, and $\support$ limits how far outside of a sphere a point may be before its contribution falls to zero.
As is typical in robust optimization (e.g. IRLS) the weights are computed with respect to the parameters $\sphere^{t-1}$ at the previous iteration -- inscription is evaluated only for points within, or in proximity of the sphere in its previous geometric configuration $\sphere^{t-1}$. 


\begin{figure}[b]
\centering
\begin{overpic} 
[width=\linewidth]
{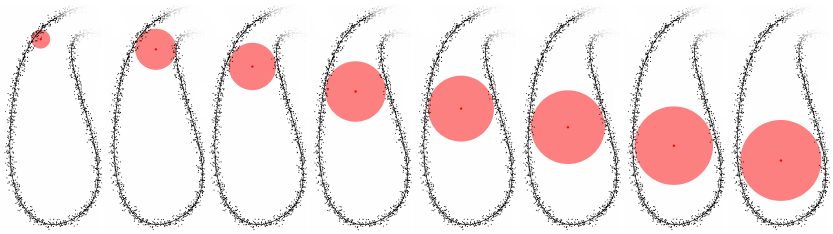}
\myfigurename{}
\end{overpic}
\caption{
While the algorithm is initialized in a neighborhood, optimizing for larger spheres will cause the solver to have the sphere converge to areas of locally maximal radius.
}
\label{fig:sliding}
\end{figure}

\paragraph{Pinned spheres}
The medial axis provides an estimate of the \emph{local thickness} of the shape through its sphere local radius.
However, as our variational formulation attempts to create \emph{larger} spheres, nothing prevents a sphere from traveling along medial branches wherever we have a non-vanishing medial radius gradient on $\mathcal{M}$; e.g. a sphere would slide from the tip of a cone to its base; see \Figure{sliding}.
We can avoid this issue by ``pinning'' medial spheres.
Generalizing the exact constraints of the sphere-shrinking algorithm by~\cite{bubble}, we subject the sphere associated with a given point $\point$ to the \emph{hard} constraint $\| \c - \point \| - \r \leq d_\text{pin}$, which keeps the sphere in contact with a sphere of radius $d_\text{pin}$ centered at $\point$. 
We include this constraint in our optimization via the penalty method~\cite{penalty}, yielding a quadratic barrier energy:
\begin{equation}
E_\text{pinning} = \ramp(\| \c - \point \| - (\r + d_\text{pin}))^2
\label{eq:pinning}
\end{equation}

\paragraph{Optimization}
Similarly to~\cite{lop}, our optimization induces the definition of medial sphere as the \emph{fixed point solution} of an update equation:
\begin{equation}
\sphere = \mathcal{F}(\sphere) = \argmin_{\sphere} \:\: E_\text{medial} + E_\text{pinning}
\label{eq:update}
\end{equation}
Our optimization problem is quadratic, but while its energy terms are differentiable, they are not linear. Hence, we iteratively compute a solution via Gauss-Newton. This requires the linearization of the arguments of the quadratic functions with respect to $\c$ and $\r$, which is straightforward in our setting.

\paragraph{Implementation}
Due to the independent nature of the per-sphere optimization, our implementation is straightforward. At each step we compute Jacobian matrices and residual vectors separately for each sphere and thus are only required to solve a $DxD$ linear system for each, where $D$ is the number of degrees of freedom of a single sphere (3 or 4 in our experiments). This property eliminates the need for any advanced solvers or factorizations and enables the implementation to be completely parallel over the spheres, as required for effective GPU acceleration. Additional performance optimization could be achieved by collecting the points for each sphere whose contribution is non-zero using accelerating data structures such as kd-trees \cite{friedman1977algorithm}.

\section{Results and evaluation}
\label{sec:eval}

We show medial representations generated by our method, in both 2D and 3D, throughout the paper and in \Figure{gallery}. As shown, our method produces valid medial representations for a wide variety of shapes and models, and in the presence of noise and outliers. Throughout the paper, we process all input with the same parameters whose values were set according to the analysis in \Section{paramanalysis}.
We evaluate our method in several ways. First, we consider variations of the inscription and maximality energy, and show how alternative formulations fail. Second, we evaluate the performance of our algorithm against ground truth on synthetic benchmarks contaminated by noise and outliers. Third, we compare our generated medial representations against the state-of-the-art in both 2D and 3D, again in the presence of noise. Finally, we provide an analysis of our algorithm parameters.

\begin{figure}[b]
\centering
\begin{overpic} 
[width=\linewidth]
{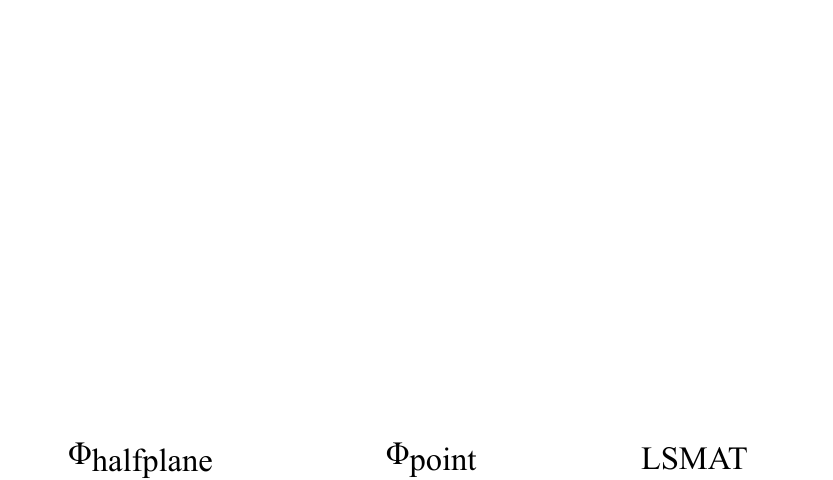}
\myfigurename{}
\end{overpic}
\caption{
Qualitative evaluation of inscription energy variants.
}
\label{fig:inscrvariants}
\end{figure}

\subsection{Variants of inscription energy -- \Figure{inscrvariants}}
We consider two modified formulations of \Equation{inscribed} in which we either penalize the squared distance to points, or the squared distance to half-planes as opposed to our blended formulation.
We investigate this behavior by randomly initializing the algorithm, and snapping $\kernel_n(\bar\c_n)$ to either zero or one for all points.
When $\kernel_n(\bar\c_n)=1$, point-to-point energy is used and the algorithm attempts to make \emph{spheres empty} in a least square sense. However, nothing prevents the optimization from generating maximal spheres \emph{outside} the shape in the ambient space.
When $\kernel_n(\bar\c_n)=1$, point-to-plane energy is used, and the algorithm attempts to make \emph{half-spaces empty}.
As illustrated in \Figure{inscrvariants} and the \textbf{supplemental video}, this results in difficulties for the algorithm in dealing with sharp concavities.
In the example above, we expect all centers to cluster to one of the two centers, but spheres in the neighborhood of the sharp concavity might read the normal of a point on the opposite side, with a halfplane requesting the radius of a sphere intersecting with it to be significantly smaller.
This problem is caused by the fact that point-to-point and point-to-plane energies approximate the squared SDF of a point set respectively in the far and near field. Our LSMAT formulation respects this geometric property, and deals with both issues at once.

\begin{figure}[t]
\centering
\begin{overpic} 
[width=\linewidth]
{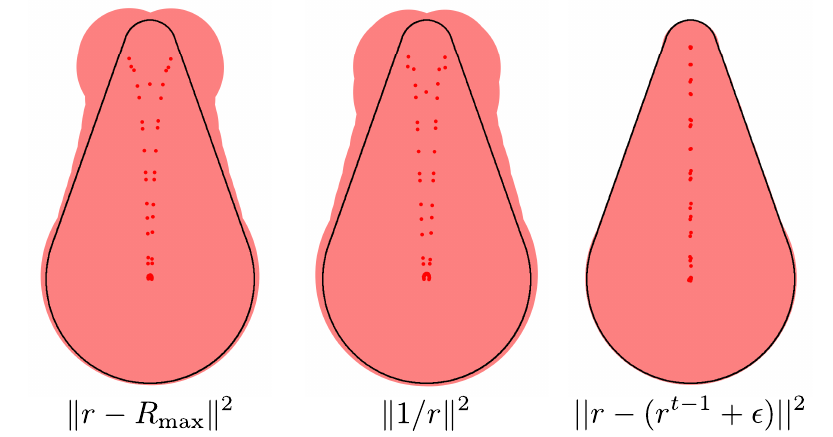}
\myfigurename{}
\end{overpic}
\caption{
Qualitative evaluation of maximality energy variants.
}
\label{fig:maxvariants}
\end{figure}

\subsection{Variants of maximality energy -- \Figure{maxvariants}}
\label{sec:eval_maximal}
We consider two alternative formulations of the maximality condition in our optimization: penalizing the squared inverse of the sphere radius, expressed as $\|1/\r\|^2$, or directly specifying $R_\text{max}$, the maximum size of a medial sphere, and optimizing $\|\r-R_\text{max}\|^2$.
While intuitively these are potentially feasible solutions, they incur a significant limitation: the amount of ``pressure'' a medial sphere will apply will be dependent on its size.
That is, small spheres will be associated with a higher energy level.
For example, using the first formulation, as $\matradii \rightarrow 0$ its gradient will tend to $-\infty$; as illustrated in \Figure{maxvariants}, this can cause spheres in the neighborhood of small features to bulge out.
Our solution, detailed in \Equation{maximal}, does not encounter this problem, as our energy is \emph{constant} regardless of the sphere size.

\begin{figure}[b]
\centering
\begin{overpic} 
[width=\linewidth]
{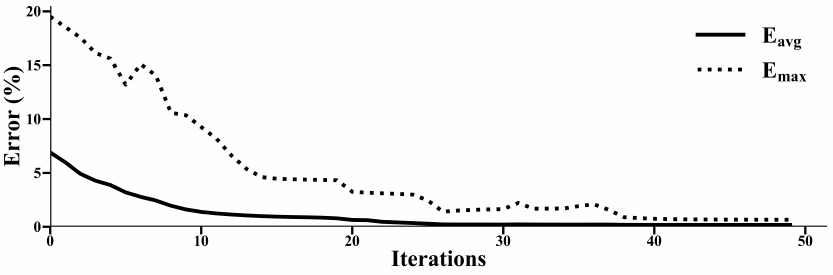}
\myfigurename{}
\end{overpic}
\caption{
As the optimization of \Equation{update} is executed, the average/max ground truth errors converge to the maximum precision.
This plot illustrates the error for the iterations visualized in \Figure{iterations}.
}
\label{fig:iterations-metric}
\end{figure}

\subsection{Quantitative evaluation metric -- \Figure{iterations-metric}}
\label{sec:metric}
We consider ground truth geometry polluted by noise, and evaluate the quality of an extracted axis by computing the distance of each medial center to the closest point on the ground truth axis~$\groundtruth{M}$:
\begin{align}
E_\text{avg} = \tfrac{1}{N} \sum_{n} \: \argmin_{\groundtruth\c_n \in \mathcal{\groundtruth{M}}} \| \c_n - \groundtruth\c_n \|_2
\\
E_\text{max} = \max_n \: \argmin_{\groundtruth\c_n \in \mathcal{\groundtruth{M}}} \| \c_n - \groundtruth\c_n \|_2 
\end{align}
We compute $\groundtruth{M}$ via the medial axis module of the python \emph{skimage} package, derived from the ridges of the distance transform from the ground truth boundary $\surface$.
As we discretize images with a bounding box at a $1024^2$ resolution, the ``numerical precision'' of the metrics above is of one pixel.
To obtain scale invariance, we report these errors in relation to the diagonal of the bounding box.
In \Figure{iterations-metric}, we visualize the convergence of our iterative optimization scheme.


\begin{figure}[t]
\centering
\begin{overpic} 
[width=\linewidth]
{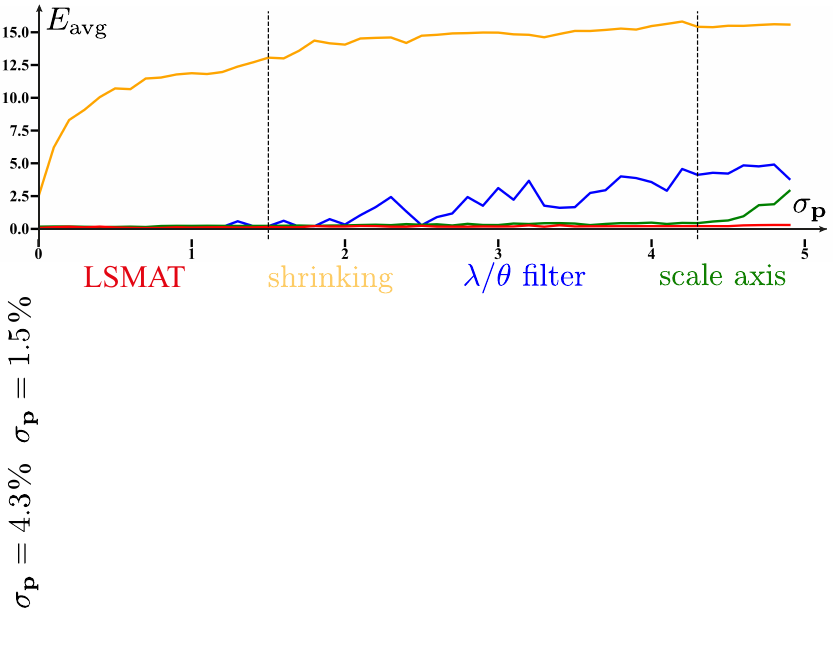}
\myfigurename{}
\end{overpic}
\caption{
(top) Quantitative evaluation on state-of-the-art methods showing increasing error as we increase the noise level. (bottom) Qualitative results corresponding to the two dashed lines in the plot.
}
\label{fig:star2d}
\end{figure}

\subsection{State-of-the-art comparisons in 2D -- \Figure{star2d}}
Through the metric from \Section{metric}, we quantitatively evaluate the performance of LSMAT against local filtering methods, as well as the global scale axis transform.
\Figure{stability} shows how neither distance nor angle filtering is very effective; hence we employ a compound variant where we first filter by distance with $\lambda>\sigma_p$, and then by angle $\theta>110^\circ$.
For the scale axis, we set $\sigma=1.3$, and for our method, we use our default parameters.
We gradually increment the level of noise, and plot the corresponding error metric in~\Figure{star2d}, as well as a few example frames from the plot. Note that LSMAT correctly captures the shape of the ground truth boundary, whereas all other local methods fail.

\begin{figure*}[t]
\centering
\begin{overpic} 
[width=\linewidth]
{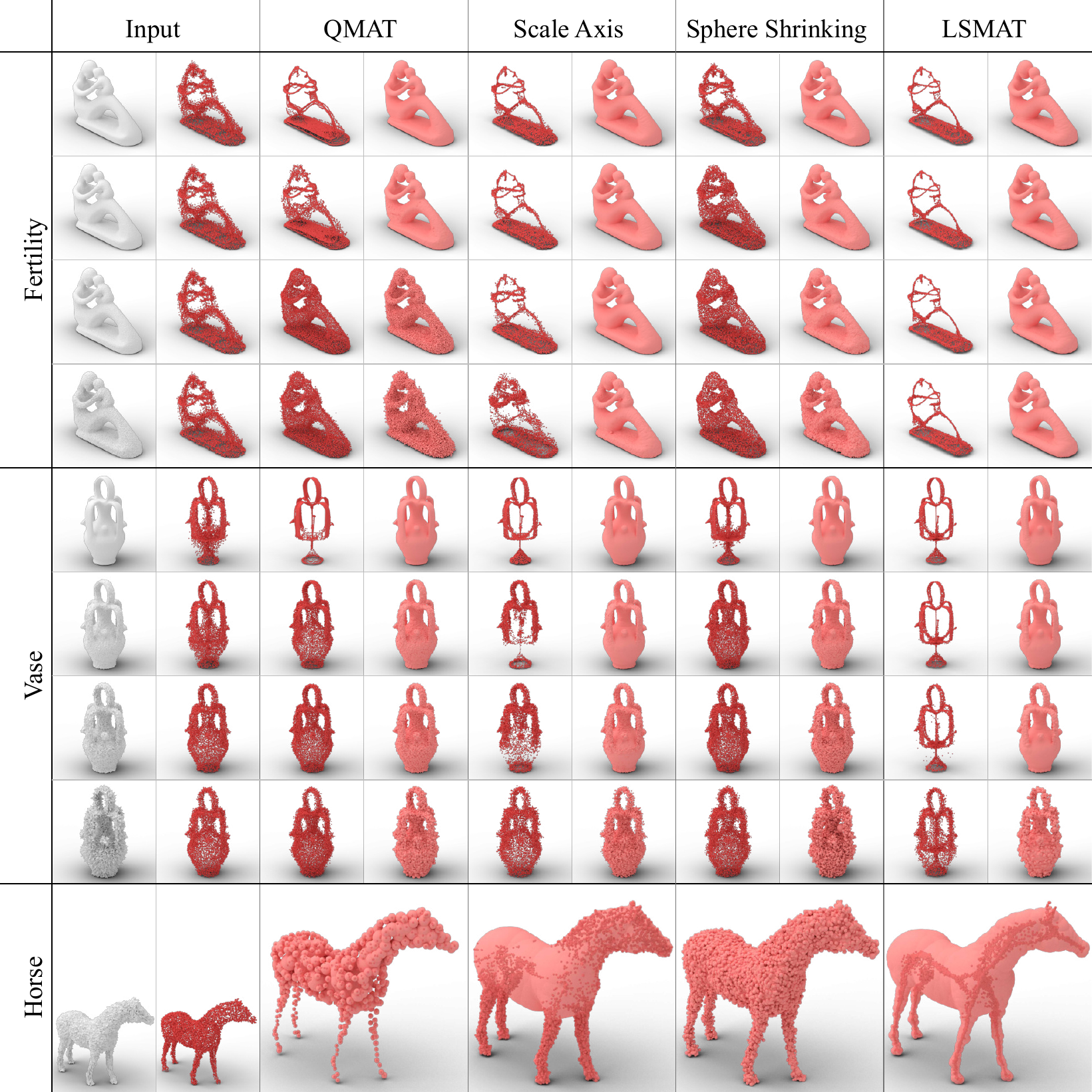}
\myfigurename{}
\end{overpic}
\caption{
Qualitative evaluation on state-of-the-art methods as we increase the noise level. The QMAT~\cite{qmat} and SAT~\cite{scaleaxis3d} in the first two columns are global methods that assume a \emph{watertight} surface, while the \bubble{}~\cite{bubble} and the LSMAT proposed here are \emph{local} methods.
}
\label{fig:star3d}
\end{figure*}

\begin{figure*}[t]
\centering
\begin{overpic} 
[width=\linewidth]
{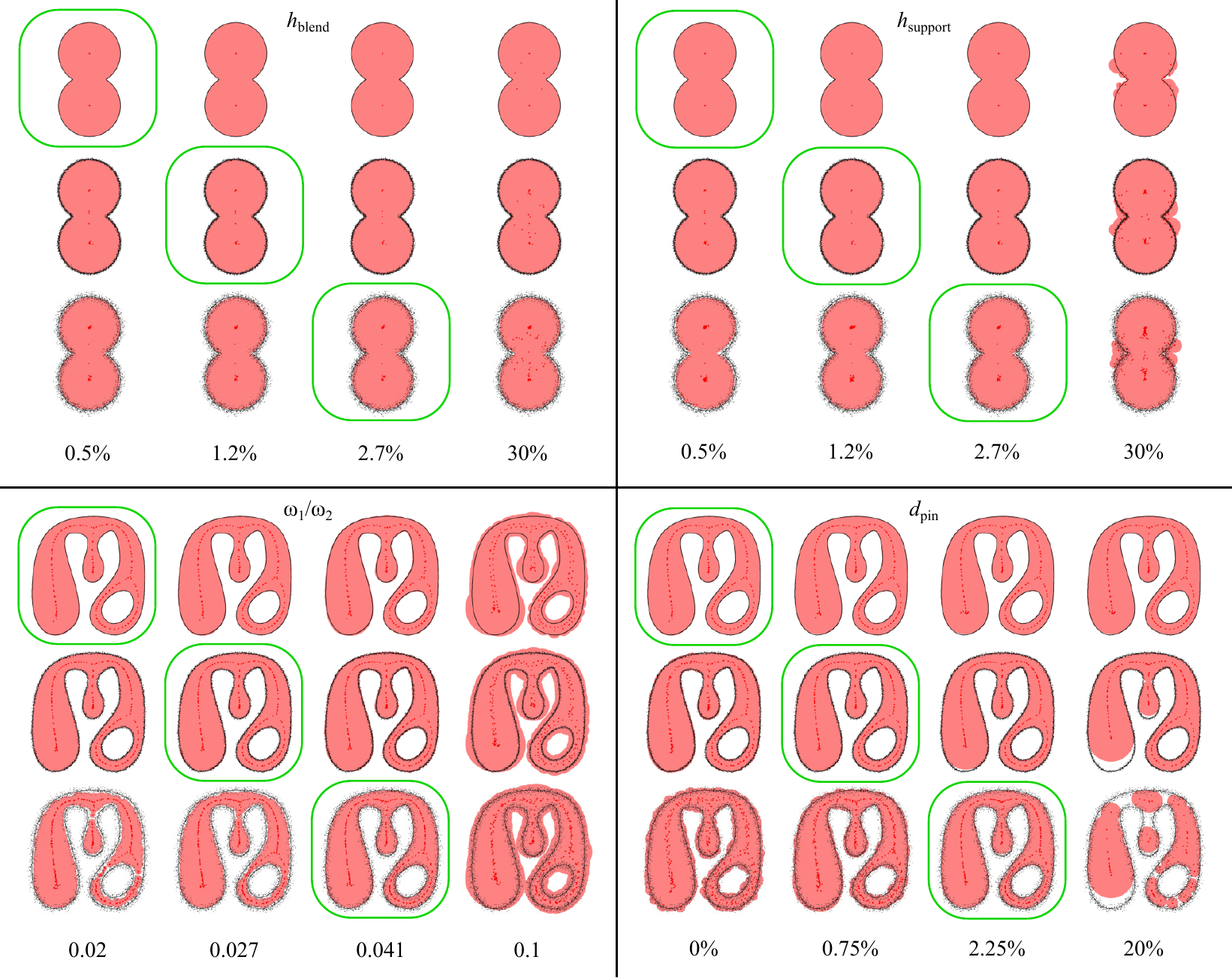}
\myfigurename{}
\put(-2,7){\vertical{$\sigma_\mathbf{p} = 3\%$}}
\put(101,7){\vertical{$\sigma_\mathbf{p} = 3\%$}}

\put(-2,18){\vertical{$\sigma_\mathbf{p} = 1\%$}}
\put(101,18){\vertical{$\sigma_\mathbf{p} = 1\%$}}

\put(-2,29){\vertical{$\sigma_\mathbf{p} = 0\%$}}
\put(101,29){\vertical{$\sigma_\mathbf{p} = 0\%$}}

\put(-2,47){\vertical{$\sigma_\mathbf{p} = 3\%$}}
\put(101,47){\vertical{$\sigma_\mathbf{p} = 3\%$}}

\put(-2,58){\vertical{$\sigma_\mathbf{p} = 1\%$}}
\put(101,58){\vertical{$\sigma_\mathbf{p} = 1\%$}}

\put(-2,69){\vertical{$\sigma_\mathbf{p} = 0\%$}}
\put(101,69){\vertical{$\sigma_\mathbf{p} = 0\%$}}
\end{overpic}
\caption{
A qualitative analysis of parameters in our algorithm. Each quadrant shows the result of sweeping a parameter (horizontal) for different noise values (vertical).
The highlighted images represent the "default" parameter choice as defined in \Section{paramanalysis}. The right column in each quadrant shows an unreasonably high choice to demonstrate that there is an upper bound for each parameter.
}
\label{fig:qualparsweep}
\end{figure*}

\subsection{State-of-the-art Comparisons in 3D -- \Figure{star3d}}
We qualitatively evaluate the performance of LSMAT in three dimensions by comparing our results to those generated by the SAT~\cite{scaleaxis3d}, QMAT~\cite{qmat}, and \bubble{}~\cite{bubble} methods.
To compare with methods that expect a mesh in input, we finely re-triangulate the surface, and apply normal displacement perturbation with $\sigma_\point \in [0\%, 2\%]$ relative to the diagonal bounding box.
On the \emph{fertility} model, LSMAT produces a convincing medial axis representation even when the input oriented point cloud is affected by extreme noise -- i.e. with a magnitude close to the one of the \emph{local feature size}.
For the \emph{vase} model, LSMAT still produces a smoother and more faithful medial axis representation than the existing state-of-the-art. 
Even in the presence of minor amounts of noise (second row), LSMAT faithfully produces a medial axis representation with a smooth surface, and whose skeleton resembles that of the uncorrupted model.
The results are particularly encouraging on the horse dataset, where we attempted to use a very small number of target primitives for QMAT ($\approx 500$) and a large value of scale for SAT ($\approx 1.3$).
Our formulation is the only one capable of computing a relatively noiseless arrangement of medial spheres. The timings for our LSMAT and sphere shrinking results can be found in \Table{timings}. All experiments were run on a machine with an Intel Xeon E5-1650 CPU and an Nvidia GTX 1080 GPU.

\begin{table}[b]
\centering
\begin{tabular}{c|c c|c c}
Model ($\sigma_\point$) & \multicolumn{2}{c|}{\textbf{LSMAT}} & \multicolumn{2}{c}{\textbf{Sphere Shrinking}} \\
10k Spheres & Time & Iterations & Time & Iterations \\
\hline
Fertility ($0\%$) & 11.8s & 40 & 0.30s &  9 \\
Fertility ($1\%$) & 11.6s & 40 & 0.35s & 11 \\
Fertility ($2\%$) & 12.0s & 40 & 0.33s & 10 \\
Fertility ($5\%$) & 16.2s & 40 & 0.34s & 10 \\
Vase      ($0\%$) & 11.2s & 70 & 0.16s &  9 \\
Vase      ($1\%$) & 12.1s & 70 & 0.17s & 10 \\
Vase      ($2\%$) & 11.8s & 70 & 0.16s &  9 \\
Vase      ($5\%$) &  6.1s & 70 & 0.16s &  9 \\
Horse     ($2\%$) &  6.8s & 60 & 0.16s &  9 \\

\end{tabular}
\caption{
Run-times for LSMAT and Sphere Shrinking results presented in \Figure{star3d}. Both algorithms are GPU accelerated and run on the same hardware.
Note that our efficiency claims are made with respect to QMAT and the scale axis transform, for which GPU acceleration is not availible.
}
\label{tab:timings}
\end{table}


\subsection{Algorithm Parameter Analysis -- \Figure{qualparsweep}}
\label{sec:paramanalysis}
Our algorithm depends on five parameters: the kernel sizes $\blend$ and $\support$, the relative weight $\omega_1/\omega_2$, pinning distance $d_\text{pin}$, and radius expansion constant $\epsilon$. The effect of varying the first four given different levels of noise is shown in \Figure{qualparsweep}. 
Note we only consider the ratio between the $\omega_*$, as the two energies balance each other. 
As $\blend$ increases, spheres near sharp concave corners begin to shrink as they eventually use the point-to-plane distance for all points in their neighborhood.
When $\support$ grows significantly beyond the local feature size, the MLS formulation is no longer able to recover a meaningful surface.
As $\omega_1/\omega_2$ increases and the relative importance of maximality versus inscription increases, spheres expand until they no longer form a faithful medial representation and ultimately escape the point cloud.
Finally, as $d_\text{pin}$ grows, the spheres are allowed to slide further from their starting positions towards areas of locally maximal radius.
We assume that estimates of the input noise characteristics are available, and choose our default value for each of these four parameters based on empirically derived linear functions of $\sigma_\point$; see Appendix \ref{paramdefaults} for details. 
Through experiments we observed that it is sufficient to choose a constant value for the fifth parameter $\epsilon$, as different values of it merely change the optimal relation between $\omega_1/\omega_2$ and $\sigma_\point$. For all our experiments we use $\epsilon = 100\%$.
We believe improved formulations of LSMAT could eventually coalesce some of these parameters hence improving the ease of use of our algorithm.

\section{Future works}
\begin{figure*}[t]
\centering
\begin{overpic} 
[width=\linewidth]
{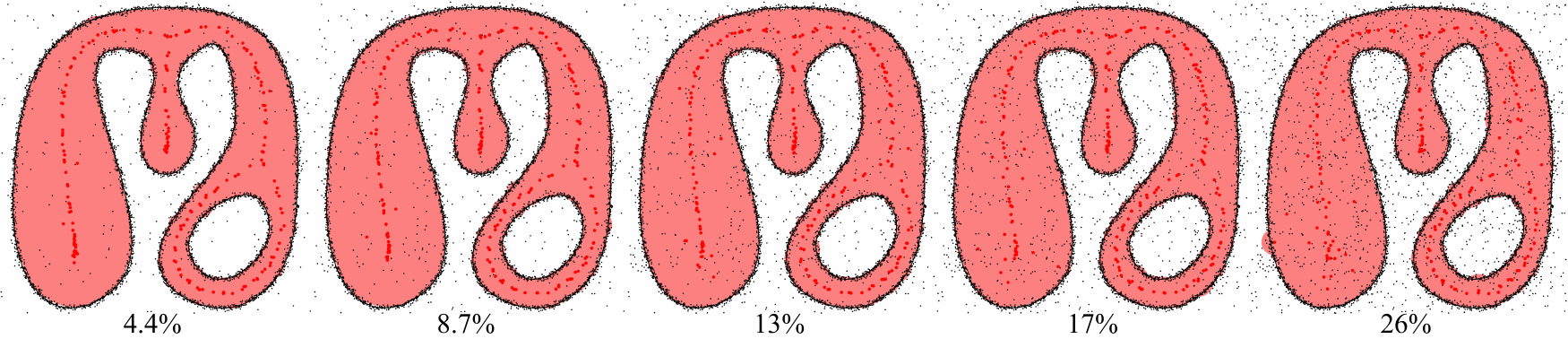}
\myfigurename{}
\end{overpic}
\caption{
Iteratively Reweighted LSMAT and its ability to cope with an increasing number of outliers (as \% of input point set).
}
\label{fig:outliers}
\end{figure*}

Expressing the Medial Axis Transform as a non-linear least squares problem opens up several interesting avenues for future research.

\paragraph{Robustness to outliers -- \Figure{outliers}}
Least squares problems can be interpreted as a maximum-likelihood (ML) estimation given a \emph{Gaussian} probability distribution of the noise variables.
If the input is corrupted by other forms of noise, one could replace \emph{Gaussian} with other error distributions, and derive the corresponding ML optimization scheme.
For example, if we assume the probability distribution of the noise to be Laplacian, the least-squares problems would simply be transformed into an $\ell^1$ (i.e. least norm) optimization.
However, these type of problems can still be computed with Gauss-Newton type methods by using iteratively re-weighted least squares~(IRLS) techniques~\cite{irls}.
As illustrated in \Figure{outliers} this results in an IR-LSMAT algorithm that can cope with significant amounts of outliers.
While these results are promising, the convergence speed of the optimization is severely reduced, as a much smaller value for $\varepsilon$ was needed to produce these results.
The generalizability of LSMAT to $\ell^p$ robust norms~\cite{sparseicp} is particularly interesting. More specifically, consider the optimization problem consisting only of the following energy:
\begin{equation}
\argmin_{\c,\r} \:\: \sum_n |\|\c-\point_n\| -r |^{p}.
\label{eq:sparsemat}	
\end{equation}
For $p=2$ this simplified version of the sphere-fitting problem is convex, hence generating a single solution regardless of initialization.
However, as $p \rightarrow 0$ the problem is non-convex and the local minima reached by optimization depends on the initialization.
Our preliminary investigation revealed how these local minima correspond to spheres belonging to the \emph{symmetry set}, a superset of the medial axis; see \Figure{definitions}(d) and \cite[Sec.~2.1.4]{skelstar}.
How to exploit \eq{sparsemat} to efficiently compute the MAT of a point set is an interesting venue for future works.

\begin{figure}[b]
\centering
\begin{overpic} 
[width=\linewidth]
{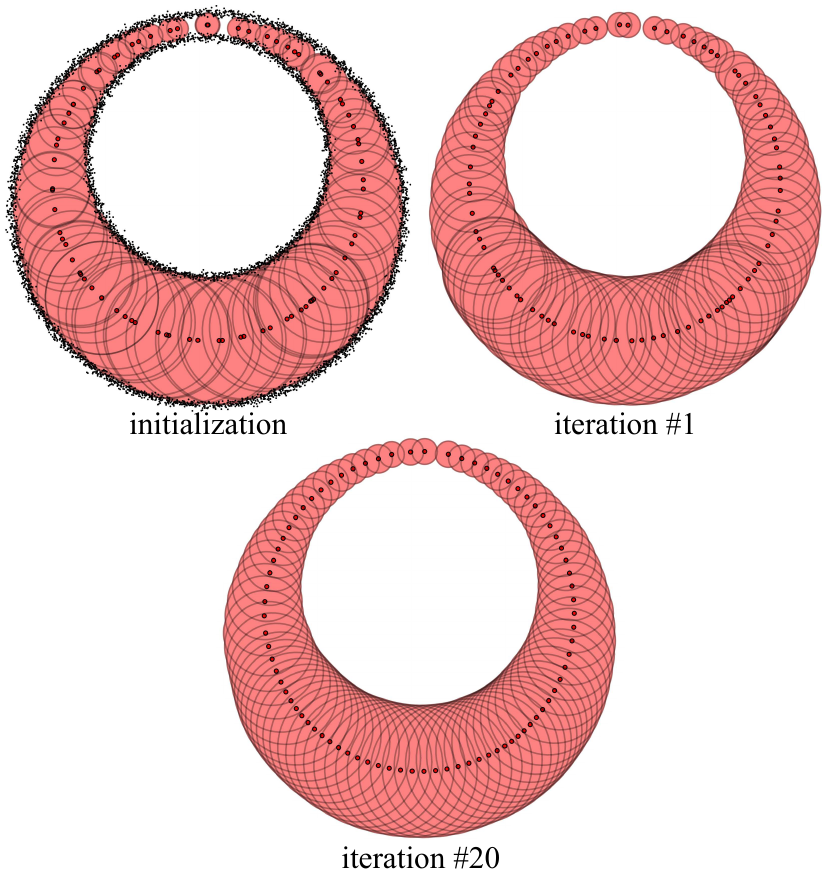}
\myfigurename{}
\end{overpic}
\caption{
Optimizing LSMAT centers placement. To highlight the smoothness of the resulting shape, we only display the input point set overlaid to the initialization.
}
\label{fig:smoothing}
\end{figure}

\paragraph{Smoothness priors -- \Figure{smoothing}}
\label{sec:postprocess}
Our pinning formulation is highly efficient, but its local nature can be also considered a intrinsic limitation.
For example, in the noisy \emph{maple leaf} example in \Figure{gallery}, at times the estimated medial spheres could be over and/or under-estimated in size, resulting in medial centers that do not necessarily sample the underlying piecewise-smooth manifold of the MAT.
However, if our input is a sampling of a smooth surface $\surface$, then we know that~\cite{matbook}: \circledone{} the medial centers $\c_*$ should be lying on a piecewise smooth manifold, and \circledtwo{} the sphere radius function on this manifold should vary smoothly.
Another desirable characteristic might be to have a uniform sampling of this manifold.
We can convert these priors into least-squares energies, resulting in a \emph{maximum a-posteriori} optimization.
In \Figure{smoothing} we illustrate a few iterations of this optimization on the \emph{moon} shape, where the ``pinning'' constraints from \Section{optcore} have been disabled.
While the optimization behaves as expected, this suffers similar shortcomings to those illustrated in \Figure{sliding}, and would result in a single sphere if executed for $t \rightarrow \infty$.
Modifying the variational LSMAT formulation to obtain a regularly sampled distribution of medial centers is an interesting venue for future works.

\paragraph{Optimization acceleration}
In our experiments, we initialize the optimization with random sphere positions and radii.
Nonetheless, given how a smooth object is composed of smooth piecewise manifolds $\axis$, and a smooth radius function $\radfun$ defined thereon, one could easily envision a locally bootstrapped version of the algorithm, where unsolved medial balls are initialized with the $\c_*,\r_*$ of their neighbors -- which could also be re-interpreted as a \emph{multi-scale} solver. 
Notice that techniques such as the \bubble{} algorithm from \cite{bubble} do not permit this type of acceleration.   

\paragraph{Shape approximation vs. reconstruction}
A number of methods leverage the medial axis for \emph{interpolatory} reconstruction~\cite{ireconstar}, and our work is a stepping stone towards the creation of \emph{approximating} reconstruction~\cite{reconstar} algorithms based on the medial axis.
Methods such as Medial~Meshes~\cite{medialmesh} and QMAT~\cite{qmat} assume a reconstruction of the watertight surface is \emph{already} available, and attempt to extract its \emph{approximation} via swept-spheres.
Conversely, our work could be extended to directly compute a \emph{reconstruction} of the input point cloud by minimizing data fitting metrics based on Hausdorff distances~\cite{medialmesh} or spherical quadrics~\cite{qmat}.
A ``topology surgery'' step could then be interweaved with our optimization to stitch medial spheres together and create a medial mesh with connectivity information.

\paragraph{Orienting a point set}
Finally, while our formulation is based on an oriented point set and an approximation of its SDF in the near/far field, an interesting variant of our algorithm could consider an unoriented point set, where the quantities to be optimized for would be the radii $[\r_1, \r_2]$, and contact plane $[\mathbf{k},\mathbf{n}]$ of a pair of \emph{twin spheres}.
The contact point $\mathbf{t}$ should then be then be optimized on the manifold $\surface$, while the complementary outside/inside label of each sphere be optimized to result in a \emph{smooth signing} of the environment space.
This approach would then provide a ``medial axis'' analogous to recent efforts in variational reconstruction of non-oriented point clouds~\cite{signing}.

\section{Conclusions}
\label{sec:conclusions}
We have introduced the Least Squares Medial Axis Transform, or LSMAT, a continuous relaxation of the medial axis transform that is not only stable, but also robust to high levels of noise even though it is based solely on \emph{local} optimization.
While in most of the paper we visualized the generated maximal spheres covering more or less the entire shape, we would like to remind the reader that the algorithm operates on each sphere \emph{independently}; the algorithm is therefore trivially parallelizable and particularly suitable for GPU implementations.
Our method produces results on noisy inputs that state-of-the-art methods fail to handle, without a reliance on ad-hoc postprocessing.
Our approach is efficient, parallelizable, and therefore suitable for real time applications where reliable medial representations are required, and where captured inputs are likely to be noisy.

\section*{Acknowledgements}
We would like to express our gratitude to Mathieu Desbrun, Leonidas Guibas, Justin Solomon, Keenan Crane, and especially Sofien Bouaziz for their ideas and suggestions.
In our experiments we use models provided courtesy of \cite{chen2009benchmark} and the AIM@SHAPE Shape Repository.

\bibliographystyle{eg-alpha}
\bibliography{lsmat}

\appendix
\section{Gradients of Energy Components}
Gradients are written in the form $\nabla_\sphere f(\sphere) = \left[\nabla_\c f(\sphere), \frac{\partial f(\sphere)}{\partial \r} \right]$.
\\
\paragraph{Inscription Term}
Given oriented surface point $(\point_n, \normal_n)$:
\begin{align}
\nabla_\sphere \sdf_\text{plane}(\sphere) &= \heaviside(\sdf_\text{plane}(\sphere)) [-\normal_n, 1]
\\
\nabla_\sphere \sdf_\text{point}(\sphere) &= \heaviside(\sdf_\text{point}(\sphere)) \left[\frac{\point_n - \c}{\|\point_n - \c\|_2}, 1 \right]
\end{align}
\paragraph{Maximality Term}
\begin{align}
\mathbf{R}_\text{maximal} &= \r - (\r^{t-1} + \epsilon)
\\
\nabla_\sphere \mathbf{R}_\text{maximal} &= \left[0, 1 \right]
\end{align}
\paragraph{Pinning Term}
Given pin point $\point$:
\begin{align}
\mathbf{R}_\text{pinning} &= \ramp(\| \c - \point \| - (\r + d_\text{pin}))
\\ 
\nabla_\sphere \mathbf{R}_\text{pinning} &= \heaviside(\mathbf{R}_\text{pinning}) \left[\frac{\c - \point}{\|\c - \point\|_2}, -1 \right]
\end{align}
\section{Empirical Parameter Defaults}
\label{paramdefaults}
Default value of $\omega_1/\omega_2$:
\begin{align}
\omega_1/\omega_2 &= 0.007 \sigma_\point + 0.02
\end{align}
Default value of $\blend$ and $\support$ (both the same):
\begin{align}
\blend &= 0.74 \sigma_\point + 0.49
\end{align}
Default value of $d_\text{pin}$:
\begin{align}
d_\text{pin} &= 0.75 \sigma_\point
\end{align}

\begin{figure*}[t]
\centering
\begin{overpic} 
[width=\linewidth]
{\currfiledir/item.png}
\myfigurename{}
\end{overpic}
\caption{
A gallery of LSMAT results with varying levels of noise on shapes with complex topology and varying feature size.
In the callout for the octopus, notice how the MLS kernel overlaps nearby surfaces, yet the algorithm can cope by producing erroneously located medial spheres with \emph{zero} radius. 
}
\label{fig:gallery}
\end{figure*}

\begin{figure*}[t]
\centering
\begin{overpic} 
[width=\linewidth]
{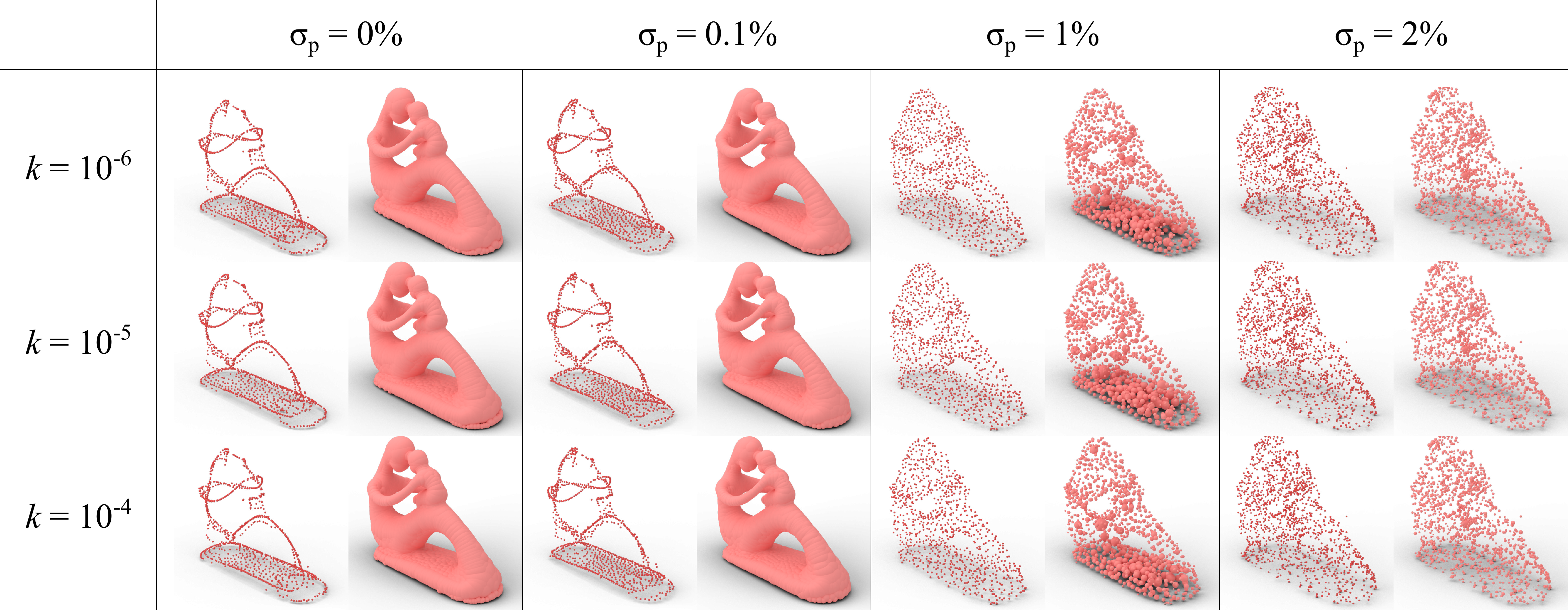}
\myfigurename{}
\end{overpic}
\caption{
Parameter sweep for QMAT. The X axis represents noise, while the Y axis shows results for different values of the parameter $k$. We found that $k$ had little effect on the noise tolerance of the algorithm. For the left two columns, the noise values are the minimum and maximum noise levels shown in the original publication, which yield good results. However, for the higher noise values that we test against QMAT fails to produce a useful medial representation. As noted in \Section{shapeapprox}, this is a known and acknowledged limitation for this family of methods.
}
\label{fig:qmatsweep}
\end{figure*}

\begin{figure*}[b]
\centering
\begin{overpic} 
[width=\linewidth]
{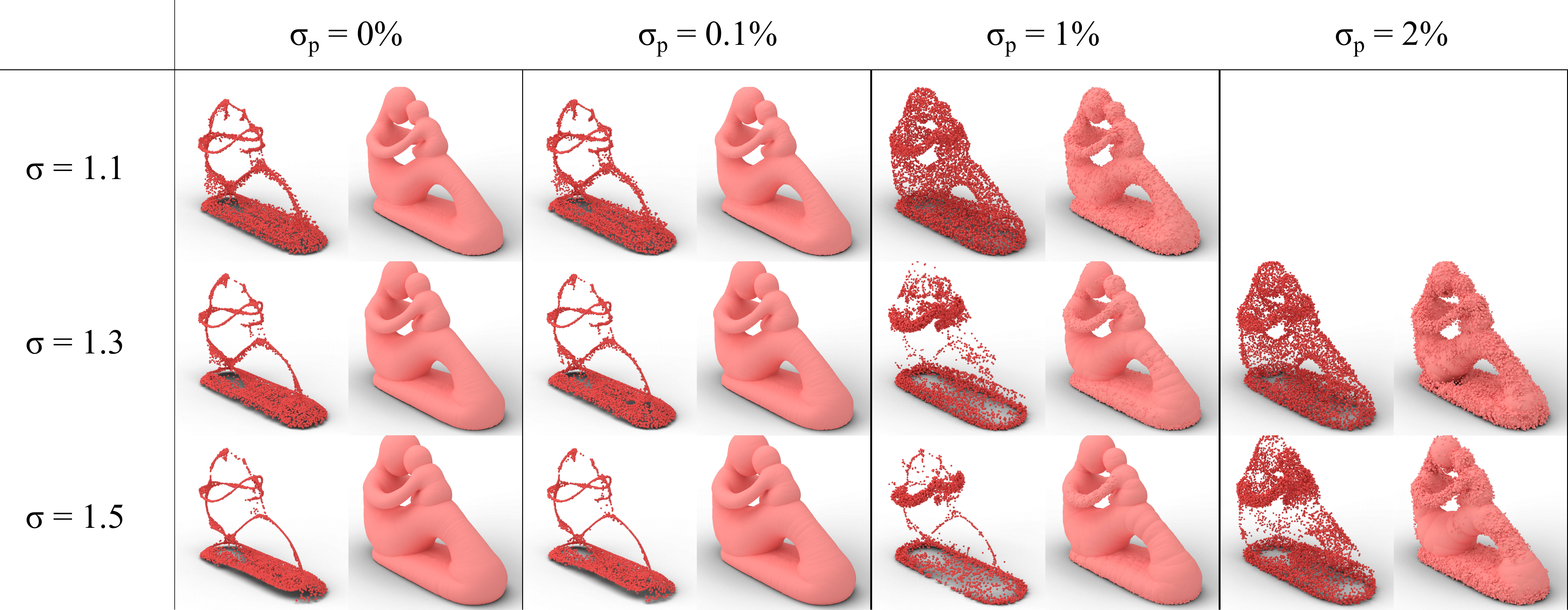}
\myfigurename{}
\end{overpic}
\caption{
Parameter sweep for Scale Axis. The X axis shows the same noise levels used in \Figure{qmatsweep}, while the Y axis shows results for different values of the scale parameter $\sigma$. As expected, this method performs well in areas where the local feature size is much larger than the noise. We show the intermediate up-scaled spheres to demonstrate this more clearly. The Scale Axis Transform is in some cases able to recover a useful medial representation even when the unfiltered MA is highly corrupted. The missing images are due to the implementation failing to complete within the allowed time for that combination of parameters.
}
\label{fig:sasweep}
\end{figure*}

\end{document}